# Antisymmetric magnetoresistance in magnetic multilayers with perpendicular anisotropy

X. M. Cheng, S. Urazhdin, O. Tchernyshyov, and C. L. Chien

*Department of Physics and Astronomy, The Johns Hopkins University, Baltimore,*

*Maryland 21218, USA*

V.I. Nikitenko, A.J. Shapiro and R.D. Shull

*National Institute of Standards and Technology, Gaithersburg, Maryland 20899, USA*

While magnetoresistance (MR) has generally been found to be symmetric in applied field in non-magnetic or magnetic metals, we have observed antisymmetric MR in Co/Pt multilayers. Simultaneous domain imaging and transport measurements show that the antisymmetric MR is due to the appearance of domain walls that run perpendicular to both the magnetization and the current, a geometry existing only in materials with perpendicular magnetic anisotropy. As a result, the extraordinary Hall effect (EHE) gives rise to circulating currents in the vicinity of the domain walls that contributes to the MR. The antisymmetric MR and EHE have been quantitatively accounted for by a theoretical model.

Several new magnetoresistance (MR) effects of both scientific and technological importance, including the giant magnetoresistance (GMR) in multilayers [1-3], the tunneling magnetoresistance (TMR) in tunnel junctions [4-6], and the colossal magnetoresistance (CMR) in perovskites [7-9], have been discovered in recent years. Other well-known MR effects include the ordinary magnetoresistance (OMR) in non-magnetic metals [10] and the anisotropic magnetoresistance (AMR) in ferromagnetic metals [11-12]. Although the mechanisms of various MR effects are different, all these intrinsic MR effects share the common symmetry in its field dependence of $\Delta R(H) = \Delta R(-H)$, i.e. they are symmetric with respect to the sign of the magnetic field $H$. In some rare cases, such as mesoscopic spin glasses, MR shows asymmetry due to frozen scattering potential of magnetic ions in mesoscopic regimes, where the time reversal symmetry is broken[13-15]. In contrast, the Hall resistance $R_H$ due to extraordinary Hall effect (EHE) [16], being proportional to the magnetization $M$, has the characteristics of $\Delta R_H(H) = -\Delta R_H(-H)$, i.e. it is antisymmetric in $H$.

In this Letter, we report the observation of antisymmetric MR with $\Delta R(H) = -\Delta R(-H)$ in multilayers with perpendicular magnetic anisotropy (PMA). By performing simultaneous magnetic domain imaging and transport measurements, we show that this new type of MR is due to the special geometry, in which the domain walls, the current and the magnetization are mutually perpendicular. The MR in this special geometry is the direct consequence of the domain structure in the multilayers with circulating currents surrounding the domain walls. Theoretical calculations show quantitative agreement with the experimental results.

The [Co/Pt]$_n$ multilayers that we have used in this work are among the few materials with well established PMA [17-22]. The Pt(100 Å)/[Co(6 Å)/Pt (10 Å)]$_n$/Pt(20 Å) multilayers with n=1, 2, 4, 8 were made by magnetron sputtering. The multilayers were lithographically patterned into Hall bars 40 μm in width and a 150 μm separation between the voltage leads for electrical transport measurements. The main results were obtained from a Pt (100 Å)/Co (3-6 Å wedge)/Pt (30 Å) trilayer, in which the crucial role of a single domain wall can be unequivocally demonstrated. The trilayer with a wedged Co layer is about 4.4 mm long along the wedge direction and 6.2 mm wide. In the wedged trilayer there is only one domain wall that separates the two macroscopic domains, and its position can be experimentally controlled. Simultaneous transport measurement and domain imaging using the magneto-optical Kerr effect (MOKE) technique [20, 23] were performed while the external magnetic field, $H$, was applied perpendicular to the film plane.

In multilayers with PMA, since $R_H$ is proportional to the magnetization component perpendicular to the film plane, $M$, the field dependence of $R_H$ is the same as that of the hysteresis loop, as confirmed by magnetometry measurements. As an example, the hysteresis loop of [Co/Pt]$_4$ multilayers, exhibiting sharp reversals, is shown in Fig. 1(a). The Hall resistance is antisymmetric with respect to the field as expected. However, the MR of the same sample, shown in Fig. 1(b), unexpectedly displays also the antisymmetry, instead of the even symmetry generally observed in most other MR. The antisymmetric MR is not due to the misalignment of the voltage leads. Indeed, when we deliberately misaligned voltage leads, there is a contribution from the EHE, but the MR with odd symmetry remains.

As shown in Fig. 1(b), the value of MR varies appreciably only during magnetization reversal, exhibiting peaks during the sharp reversal. MOKE images show that in decreasing field, the reversed domains nucleate from several nucleation centers within the field of view and expand in the form of bubble domains [darker regions in the inset of Fig. 1(b)]. In increasing field, bubble domains of opposite $M$ direction nucleate from the same nucleation centers as those in decreasing field, while a peak of opposite polarity in MR is observed. Both the evolution of the bubble domains and the resulting MR peaks are reproducible. These results indicate a direct relationship between the domain structure and the peculiar MR peaks of opposite polarity. However, the complexity of domain structure evolution in [Co/Pt]$_n$ multilayers renders it unsuitable for quantitative studies.

The specially designed Pt/wedged-Co/Pt trilayer, which is a [Co/Pt]$_n$ multilayer with n = 1, contains a wedged ferromagnetic Co layer, as schematically shown in Fig. 2(a). As confirmed by direct domain MOKE imaging, the magnetization reversal in this specially designed wedged sample involves only two macroscopic domains. The two domains, extending across the entire sample, are separated by a 180$^o$ domain wall, which runs perpendicular to the wedge direction [Fig. 2 (b)]. As the magnitude of the reverse field increases, the domain wall appears from the thin end, sweeps across the sample along the wedge direction, and finally disappears at the thick end. Upon reversing the magnetic field, new domain of opposite magnetization appears again from the thin edge and propagates towards the thick end. The domain wall that separates the two domains can be frozen in position by switching off the external field [20]. In this manner, the position of the domain wall $x_{DW}$ (relative to the center line $x_C$ of the specimen) can be located to any location along the wedge direction by using the applied magnetic field as shown in Fig. 2(c).

During the MR and the EHE measurements using the wedged sample, the current was sent from $I_1$ to $I_2$ perpendicular to the domain wall as shown in Fig. 2(a). The MR measurements were performed across $V_2$ and $V_4$ at the lower edge, and across $V_1$ and $V_3$ at the upper edge as shown in Fig. 2(a). The MR measured across $V_2$ and $V_4$ are shown in Fig. 3(a). When the wedged Co layer is in the single-domain state, a resistance value of $R_S$ is obtained. The MR measurements and MOKE images conclusively demonstrate that any resistance, appreciably different from $R_S$, is the direct result of the appearance and the propagation of a single domain wall. For the decreasing-field branch, the MR is negative with its value decreasing from $R_S$ at $-4.0$ mT when the reversed domain appears from the thinner end. It reaches the most negative value at -9.5 mT when the domain wall is midway between the two voltage contacts.

It is clear that the MR measured across $V_2$ and $V_4$, shown in Fig. 3(a), is antisymmetric in $H$, showing a negative peak and a positive peak for the decreasing and increasing field branch respectively. Equally unexpected, the MR measured across $V_1$ and $V_3$, shown in Fig. 3(b), instead of being the same as that in Fig. 3(a) as generally expected for most MR, has the *opposite* sign to that measured across $V_2$ and $V_4$. Thus, not only the MR curves show the unexpected antisymmetry in $H$, the MR measured at two structurally identical and geometrically equivalent edges are also opposite in sign.

The Hall resistances $R_H$ measured across $V_1$ and $V_2$ at the thick end and across $V_3$ and $V_4$ at the thin end of the wedge layer are shown in Fig. 3(c). The values of $R_H$ are closely correlated to those of the MR. The $R_H$ measured at the thick end (solid lines) across $V_1$ and $V_2$ and at the thin end (dotted lines) across $V_3$ and $V_4$ are different because the domain wall first appears at the thin edge and passes the leads $V_3$ and $V_4$ first. After switching off the field, thus freezing the domain wall, neither the MR nor the Hall resistance subsequently varies. This shows conclusively that the observed anomalies in MR and Hall resistance measured at different locations are static in nature, and that they are related only to the domain structure, and more specifically, the location of the single domain wall.

The MR results with the odd symmetry, closely related to the Hall resistance results, are due to the special geometry provided by the multilayers with PMA, where the current, the domain wall, and the magnetization directions are mutually perpendicular. This is schematically shown in Fig. 4(a) with one domain wall separating the two domains. Such unusual antisymmetric MR can be qualitatively understood in terms of symmetry. Consider an idealized situation when the leads possess symmetries in the $x$ and $y$ planes dissecting the sample [Fig. 4(b)]. By virtues of these symmetries, longitudinal resistance $R_{13}$ and $R_{24}$ satisfy

$$R_{13}\left[H, M(x,y)\right] \equiv \frac{V_{13}}{I} = R_{13}\left[-H, -M(-x, y)\right] \quad (x\ reflection) \qquad (1a)$$

$$R_{24}\left[H, M(x,y)\right] \equiv \frac{V_{24}}{I} = R_{13}\left[-H, -M(x, -y)\right] \quad (y\ reflection) \qquad (1b)$$

When the magnetization is uniform [upper part of Fig. 4(b)], $M(x,y)=M$, longitudinal resistance $R_{13}$ and $R_{24}$ are even functions of $H$ and $M$, as is usually observed. The presence of a domain wall changes that. The maximal effect is achieved when the domain wall is at the center of the sample, in the plane $x=0$, so that $M(x,y)=M(x)= -M(-x)$ [lower part of Fig. 4(b)]. Then, Eq. (1a) reads $R_{13}[H, M(x)]= R_{13}[-H, M(x)]$, i.e., longitudinal resistance is still even in $H$, but may contain terms that are odd in $M$. MR $\Delta R_{13} \equiv R_{13}[H, M(x)]-R_{13}[0,0]$ will be quadratic in $H$ and linear in $M$ and thus more sensitive to $M$ than to $H$. The part antisymmetric in $M$ will change sign if voltage is measured along the opposite edge:

$$\Delta R_{13}[0, M(x)] - \Delta R_{13}[0, -M(x)] = -\Delta R_{24}[0, M(x)] + \Delta R_{24}[0, -M(x)]$$

on account of Eq.(1b).

The physical origin of the antisymmetric term can be understood as follows. Charge carriers are deflected in opposite directions in the two domains of the magnetization. This induces a circulating current in the vicinity of the domain wall [Fig. 4(c)]. This current is accompanied by an electric field essentially parallel to it. The electrical field points in opposite directions along the upper and lower edges and is reversed upon magnetization reversal.

To quantitatively account for the MR and the EHE results, we have computed the Hall voltage and MR for an infinitely long film of width $a$ [Fig. 4(a)]. The current density is taken to be $\mathbf{j}=\mathbf{j}_0 + \boldsymbol{\delta j}(x)$, consisting of a uniform $\mathbf{j}_0$ throughout the slab and a nonuniform current $\boldsymbol{\delta j}$ around $x=0$ due to magnetization reversal at $x=0$ [Fig. 4(a)]. The electrical field $\mathbf{E}$ and current density $\mathbf{j}$ can be related by a resistivity tensor $\rho$ in the form $\mathbf{E}=\rho\mathbf{j}$, where $\rho$ has the form

$$\begin{pmatrix} \rho & -\rho_H \operatorname{sgn}(x) \\ \rho_H \operatorname{sgn}(x) & \rho \end{pmatrix} \qquad (2)$$

In the limit of $\rho_H \ll \rho$, the nonuniform current $\boldsymbol{\delta j}(x,y)$ induced around the domain wall [Fig. 4(b)] can be treated as a small perturbation of the average current density [24-25]. Using this model we have quantitatively calculated the Hall voltage and the MR. To first order in $\rho_H$, the Hall voltage is

$$V_H(x) = (\rho_H j_0 a) \operatorname{sgn}(x) \left( 1 - \frac{8}{\pi^2} \sum_{n=odd}^{\infty} \frac{e^{-\pi n |x|/a}}{n^2} \right) \tag{3}$$

The value of the MR depends on the locations of the electrodes. For two electrodes placed at points $x = x_C \pm b/2$ along the lower edge ($y=0$), the MR is given by the expression

$$MR = \frac{R - R_S}{R_S} = \frac{\rho_H}{\rho} \frac{a}{b} \left( \frac{4}{\pi^2} \sum_{n=odd}^{\infty} \frac{\exp\left(-\left|\frac{\pi n(x_c - b/2)}{a}\right|\right) - \operatorname{sgn}(|x_c| - b/2) \exp\left(-\left|\frac{\pi n(x_c + b/2)}{a}\right|\right)}{n^2} - \theta(b/2 - |x_c|) \right) \tag{4}$$

The calculated results can be directly compared with the experimental data. Since the values of the Hall resistance and MR depend only on the location of the single domain wall $x_{DW}$, we can directly compare the theoretical and experimental results as a function of $x_{DW}$. This is shown in Fig. 4(d-e) for the MR and the Hall resistance results, demonstrating quantitative agreement. It should be noted that the special geometry of mutually perpendicular current, domain wall, and magnetization cannot be realized in most magnetic thin films and multilayers where in-plane anisotropy generally prevails.

In conclusion, we have demonstrated a new form of MR, which is antisymmetric in $H$, in multilayers with PMA. By performing simultaneous MOKE imaging and transport measurements on samples with a controlled two-domain structure in the Pt/Co wedge/Pt trilayer, we show conclusively that the antisymmetric MR originates from the Hall fields due to EHE on either side of the domain wall. The observed MR and EHE results can be quantitatively accounted for theoretically. This rare occurrence of antisymmetric MR is due to special geometry afforded in multilayers with PMA where the magnetization vector, the current direction, and the domain wall direction are mutually perpendicular.

From the perspective of symmetry, the appearance of the domain wall permits the existence of longitudinal resistance terms which are odd in *M*.

Work supported by NSF Grant No. DMR00-80031.

# References

[1] M. N. Baibich, J. M. Broto, A. Fert, F. Nguyen Van Dau, F. Petroff, P. Eitenne, G. Creuzet, A. Friederich, and J. Chazelas, Phys. Rev. Lett. **61**, 2472 (1988).

[2] S. S. P. Parkin, N. More, and K. P. Roche, Phys. Rev. Lett. **64**, 2304 (1990).

[3] John Q. Xiao, J. S. Jiang, and C. L. Chien, Phys. Rev. Lett. **68**, 3749 (1992).

[4] M. Julliere, Phys. Lett. **54A**, 225 (1975).

[5] J. S. Moodera and L. Kinder, J. Appl. Phys. **79**, 4724 (1996).

[6] J. C. Slonczewski, Phys. Rev. B **39**, 6995 (1989).

[7] I. N. Krivorotov, K. R. Nikolaev, A. Y. Dobin, A. M. Goldman, and E. D. Dahlberg, Phys. Rev. Lett. **86**, 5779 (2001).

[8] R. von Helmolt, J. Wecker, B. Holzapfel, L. Schultz, and K. Samwer, Phys. Rev. Lett. **71**, 2331 (1993).

[9] S. Jin, T. H. Tiefel, M. Mccormack, R. A. Fastnacht, R. Ramesh, and L. H. Chen, Science **264**, 413 (1994).

[10] R. C. O'Handley, Modern Magnetic Materials Principles and Applications, Wiley-Interscience Publication, p. 590, 2000.

[11] T. R. McGuire and R. I. Potter, IEEE Trans. **MAG-11**, 1018 (1975).

[12] R. I. Potter, Phys. Rev. B **10**, 4626, (1974).

[13] P. G. N. de Vegvar, L. P. Lévy, and T. A. Fulton, Phys. Rev. Lett. **66**, 2380 (1991).

[14] P. G. N. de Vegvar and T. A. Fulton, Phys. Rev. Lett. **71**, 3537 (1993).

[15] J. Jaroszynski, J. Wróbel, G. Karczewski, T. Wojtowicz, and T. Dietl, Phys. Rev. Lett. **80**, 5635 (1998).

[16] C. L. Chien, C.R. Westgate, The Hall Effect and its Applications, Plenum Press, New York, (1980).

[17] N. C. Koon, B. T. Jonker, F. A. Volkening, J. J. Krebs, and G. A. Prinz, Phys. Rev. Lett. **59**, 2463 (1987).

[18] F. J. A. den Broeder, D. Kuiper, A. P. van de Mosselaer, and W. Hoving, Phys. Rev. Lett. **60**, 2769 (1988).

[19] B. N. Engel, C. D. England, R. A. Van Leeuwen, M. H. Wiedmann, and C. M. Falco, Phys. Rev. Lett. **67**, 1910 (1991).

[20] J. Pommier, P. Meyer, G. Penissard, J. Ferre, P. Bruno and D. Renard, Phys. Rev. Lett. **65**, 2054 (1990).

[21] O. Hellwig, A. Berger, and E. E. Fullerton, Phys. Rev. Lett. **91**, 197203 (2003).

[22] O. Hellwig, S. Maat, J. B. Kortright, and E. E. Fullerton, Phys. Rev. B **65**, 144418 (2002).

[23] V.I. Nikitenko, V.S. Gornakov, A.J. Shapiro, R.D. Shull, K. Liu, S.M. Zhou, and C. L. Chien, Phys. Rev. Lett. **84**, 765 (2000).

[24] R.T. Bate, J.C. Bell, and A.C. Beer, J. Appl. Phys. **32**, 806 (1961).

[25] D.L. Partin, M. Karnezos, L.C. deMenezes, and L. Berger, J. Appl. Phys. **45**, 1852 (1974).

Figure Captions

Fig. 1. Antisymmetric field dependences of (a) Hall resistance $R_H$ and (b) MR of a [Co (6 Å)/Pt(10 Å)]$_4$ multilayer with field applied perpendicular to the film plane. The $R_H$ and MR peaks are correlated with the domains (inset of 1b) during magnetization reversal.

Fig. 2. (a) Schematic of a Pt(100 Å)/Co(3-6 Å) wedge /Pt (30 Å) specimen with a wedged Co layer from right to left, the current leads ($I_1$ through $I_4$) and the voltage leads ($V_1$ through $V_4$). The external field is applied perpendicular to the film plane. The MOKE imaging area is shown by the rectangular frame.

(b) MOKE images of domain patterns at fields of -2.0 mT (single domain with **M** up), -7.8 mT, -9.5 mT (two domains with the domain wall inside the imaging area) and -11.0 mT (two domains with the domain wall on the left of the imaging area).

(c) Location of the domain wall measured from the center of the sample as a function of external magnetic field.

Fig. 3. MR measured at (a) the lower edge across $V_2$ and $V_4$ and (b) the upper edge across $V_1$ and $V_3$ showing opposite polarity with the same current from $I_1$ to $I_2$. The resistance has the same value $R_S$ in the single domain state, and the extreme values when the domain wall is at the center line of contacts $V_2$ and $V_4$. The shaded areas indicate opposite domains. The Hall resistance (c) measured from the left edge across $V_1$ and $V_2$ (the solid line) changes more gradually than that from the right edge across $V_3$ and $V_4$ (the dotted line).

Fig. 4. Schematic of EHE (a) in an infinite long slab with a $180^0$ domain wall at $x=0$. The reversal of **M** across the domain wall causes opposite Hall fields, resulting in an effective electrical field across the domain wall of opposite signs at upper and lower edges. The appearance of the domain wall permits the existence of longitudinal resistance terms which are odd in $M$ (b). The EHE in two-domain state results in a nonuniform current $\delta\boldsymbol{j}$ (c) around the domain wall (at $x=0$) in an infinite slab with width $a$, where the length of the arrow is scaled to the magnitude of $\delta\boldsymbol{j}$. The calculated MR (d) and Hall resistance (e) as a function of the domain wall location are shown as the solid lines, in good agreement with the experimental results.

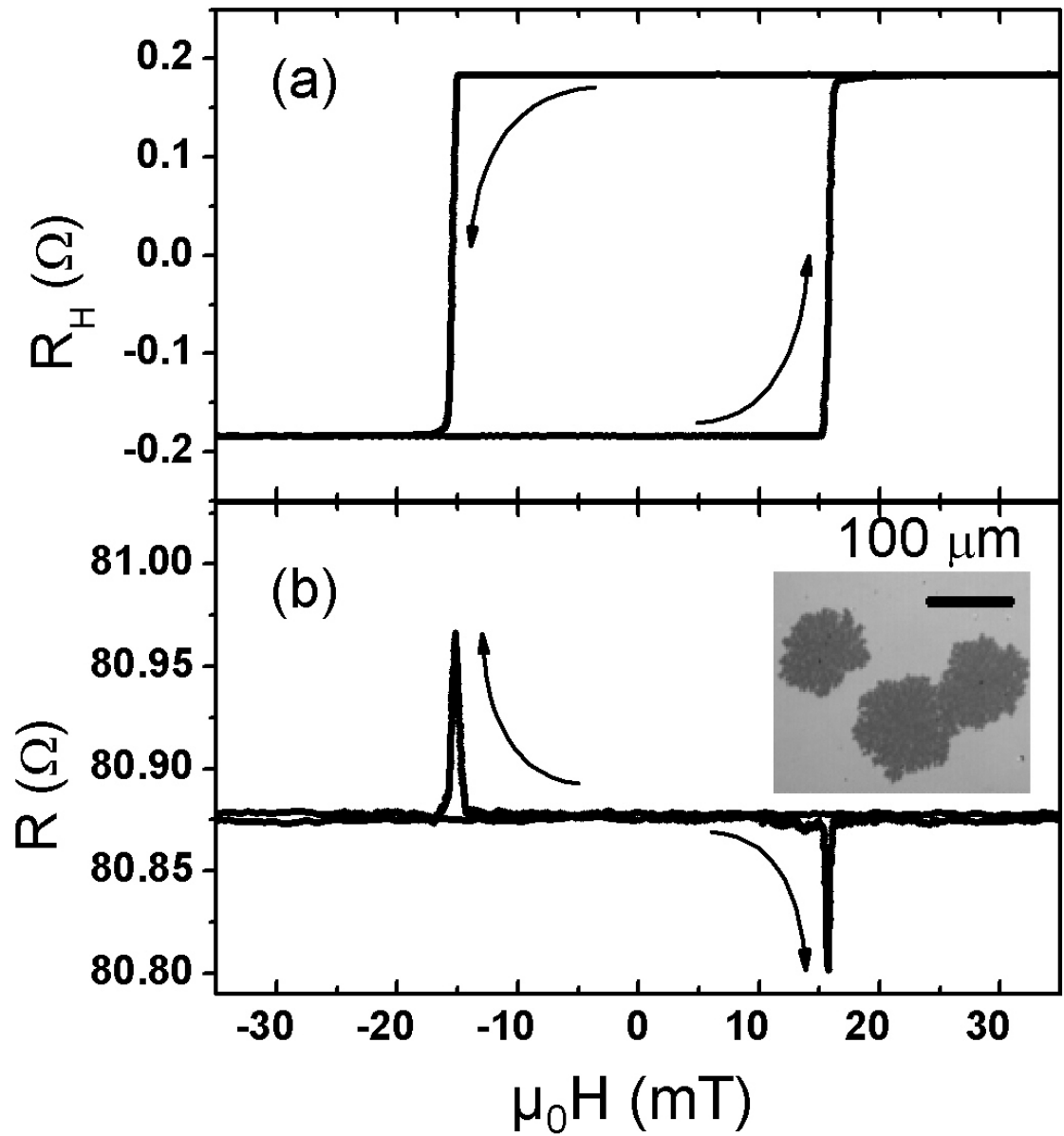

Fig. 1

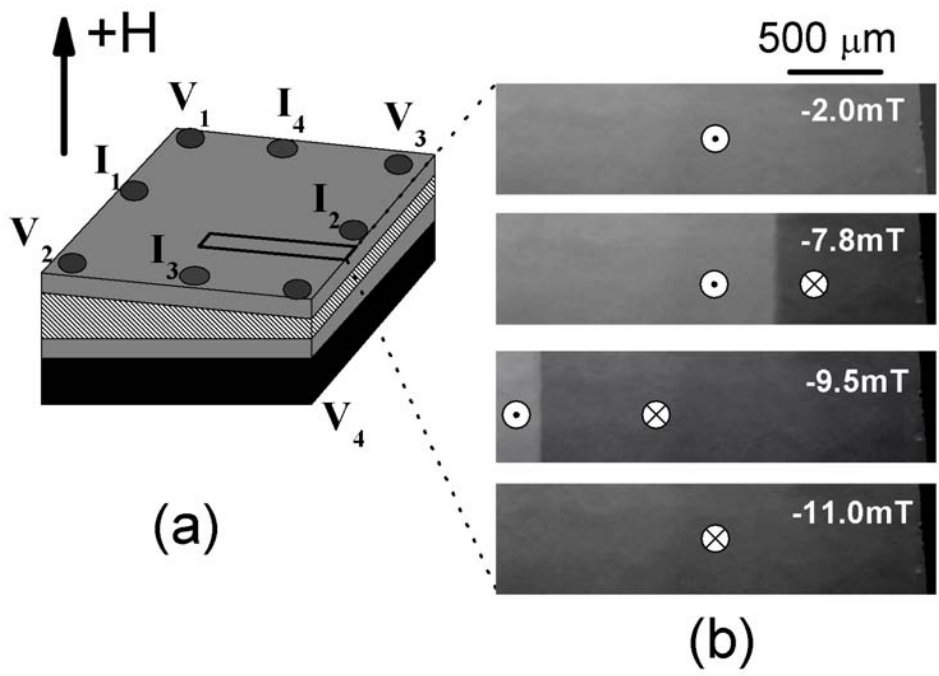

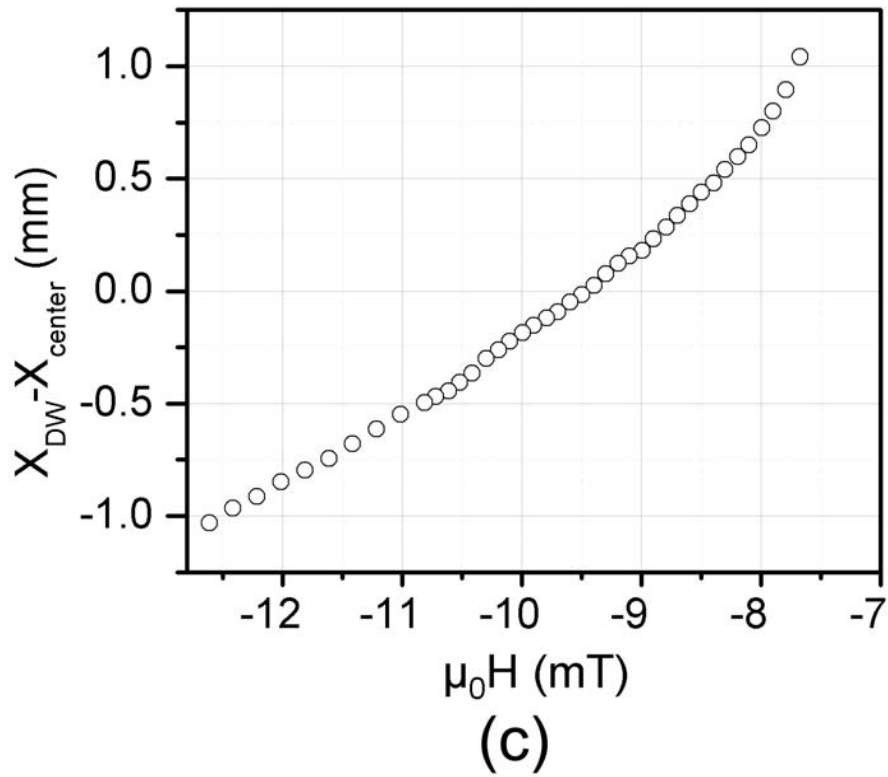

Fig. 2

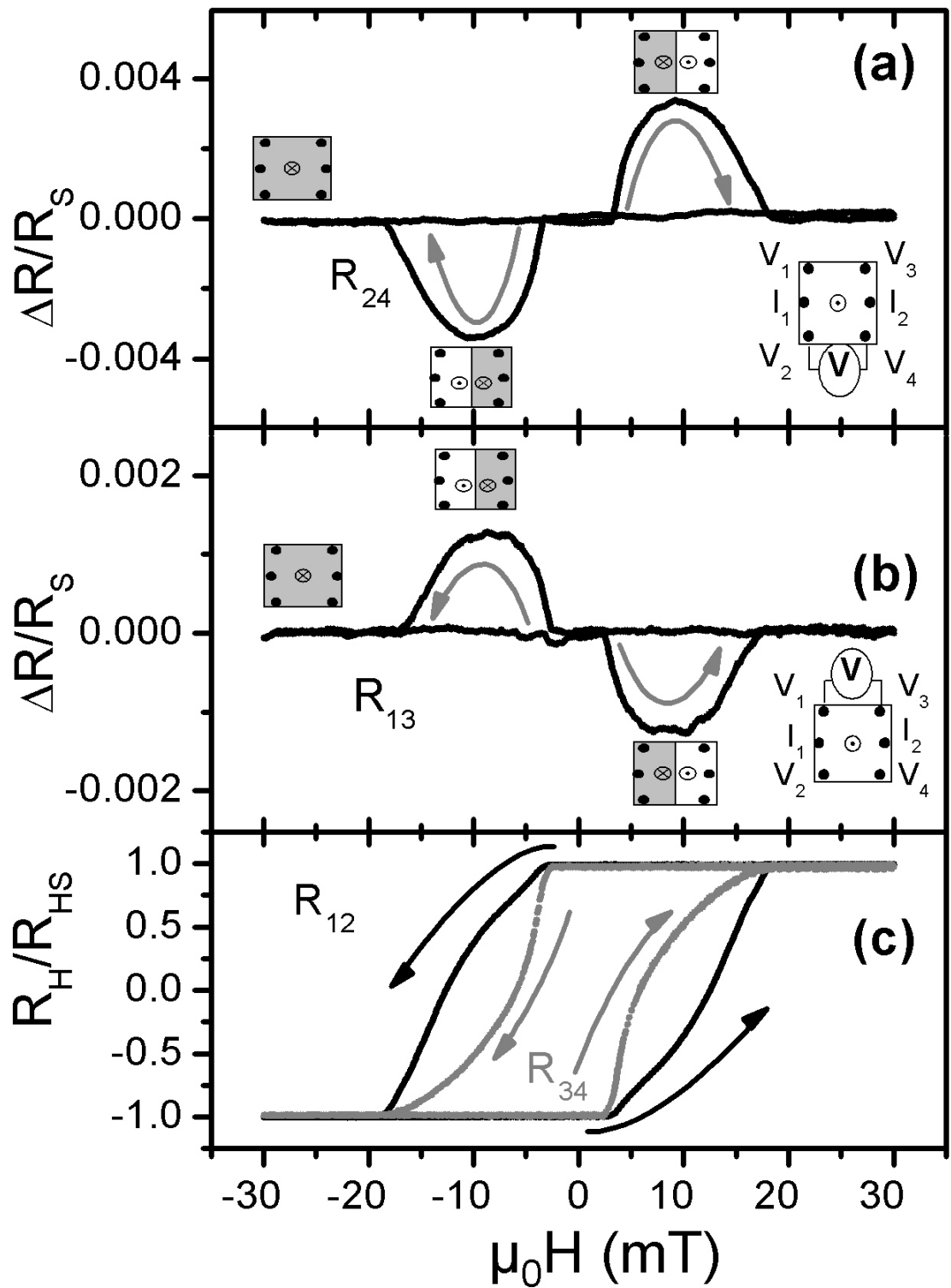

Fig. 3

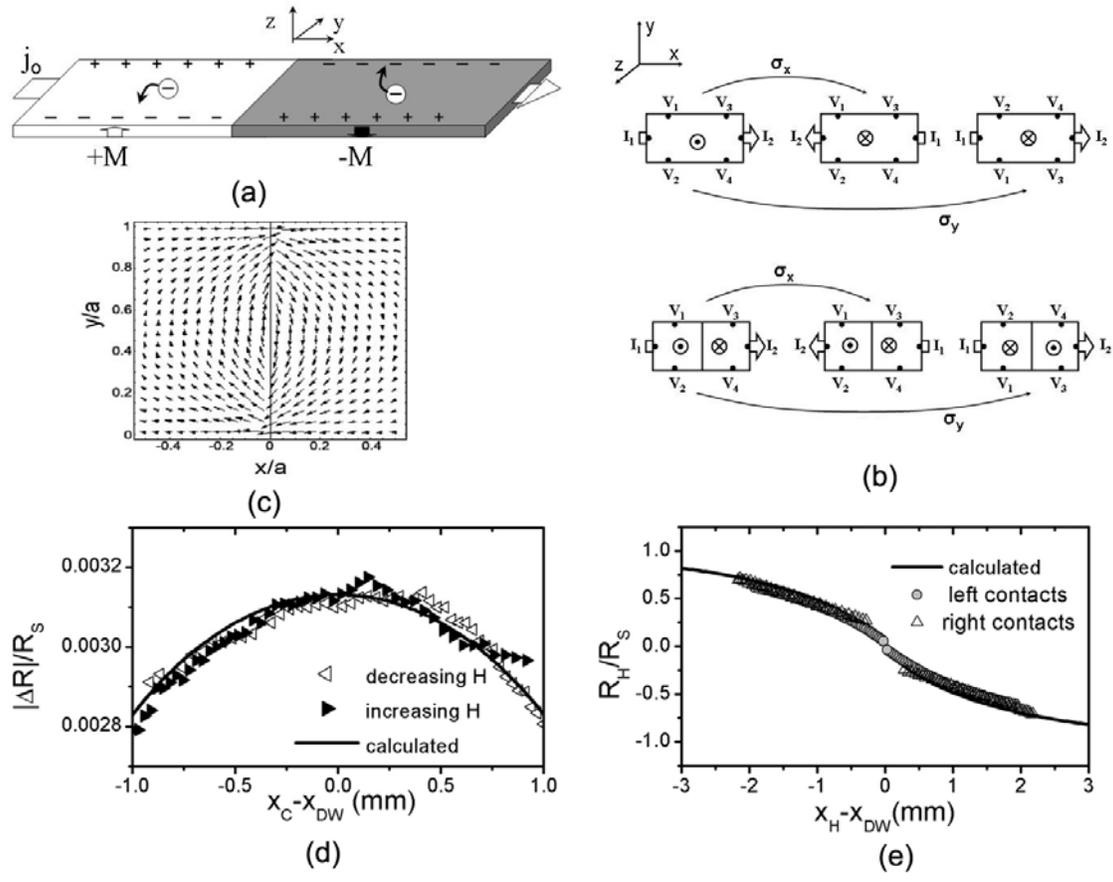

Fig. 4